\documentclass[runningheads]{llncs}
\usepackage[T1]{fontenc}

\usepackage{stmaryrd}
\usepackage{graphicx}

\usepackage{amssymb}
\usepackage{mathtools}

\usepackage{cleveref}

\usepackage[ruled,vlined,linesnumbered]{algorithm2e}
\Crefname{algocf}{Algorithm}{Procedure}
\SetAlgorithmName{Procedure}{}{}

\usepackage{tikz}
\usepackage{pgfplots}

\newcommand{\MTL}{\ensuremath{\textup{DatalogMTL}}}
\newcommand{\matA}{M}

\newcommand{\sbf}{\mathbf{s}}

\DeclareFontFamily{U}{MnSymbolC}{}
\DeclareSymbolFont{MnSyC}{U}{MnSymbolC}{m}{n}

\DeclareMathSymbol{\square}{\mathbin}{MnSyC}{106}
\DeclareMathSymbol{\meddiamond}{\mathbin}{MnSyC}{110}
\DeclareMathSymbol{\boxminus}{\mathbin}{MnSyC}{112}
\DeclareMathSymbol{\boxplus}{\mathbin}{MnSyC}{116}
\DeclareMathSymbol{\diamondminus}{\mathbin}{MnSyC}{120}
\DeclareMathSymbol{\diamondplus}{\mathbin}{MnSyC}{124}

\DeclareFontShape{U}{MnSymbolC}{m}{n}{
    <-5>  MnSymbolC4
    <5-6>  MnSymbolC5
    <6-7>  MnSymbolC6
    <7-8>  MnSymbolC7
    <8-9>  MnSymbolC8
    <9-10> MnSymbolC9
    <10-12> MnSymbolC10
    <12->   MnSymbolC12}{}

\newcommand{\Si}{\mathcal{S}}
\newcommand{\Ui}{\mathcal{U}}

\newcommand{\I}{\mathfrak{I}}
\newcommand{\D}{\mathcal{D}}
\newcommand{\ins}[2]{\ensuremath{\mathsf{inst}_{#1}[#2]}}
\newcommand{\der}[2]{\ensuremath{#1[#2]}}
\newcommand{\semi}[3]{\ensuremath{#1[ #2 \threedotcolon #3  ]}}
\newcommand{\prognr}{\Prog_{nr}}

\newcommand{\can}[2]{\mathfrak{C}_{#1,#2}}
\newcommand{\ground}[2]{\mathsf{ground}(#1,#2)}
\newcommand{\Prog}{\Pi}

\newcommand{\N}{\mathbb{N}}

\newcommand{\threedotcolon}{\,\substack{\cdot\\[-0.12cm]\cdot\\[-0.12cm]\cdot\\[0.05cm]}\,}

\newcommand{\subtag}[1]{\tag{#1}}

\begin{document}
\title{Semina\"ive Materialisation in DatalogMTL}


\author{Dingmin Wang \and Przemysław Andrzej Wałęga \and Bernardo Cuenca Grau}
\authorrunning{D. M. Wang et al.}
\institute{Department of Computer Science, University of Oxford, UK\\
\email{\{dingmin.wang, przemyslaw.walega, bernardo.cuenca.grau\}@cs.ox.ac.uk}}

\maketitle

\begin{abstract}
DatalogMTL is an extension of Datalog with metric temporal operators that has found applications in temporal ontology-based data access and query answering, as well as in stream reasoning. 
Practical algorithms for DatalogMTL are reliant on materialisation-based reasoning, where temporal facts are derived in a forward chaining manner in successive rounds of rule applications. Current materialisation-based procedures are, however, based on a na\"ive evaluation strategy, where the main source of inefficiency stems from redundant computations.  
In this paper, we propose a materialisation-based procedure which, analogously to the classical semina\"ive algorithm in Datalog, aims at minimising redundant computation by ensuring that
each temporal rule instance is considered
at most once during the execution of the algorithm.
Our experiments show that 
our optimised semina\"ive strategy for DatalogMTL is able to 
significantly reduce materialisation times.


\end{abstract}

\section{Introduction}

DatalogMTL is a temporal rule-based
language that
has found a growing number of applications 
in ontology-based data access \cite{kikot2018data,guzel2018ontop,koopmann2019ontology} and stream
reasoning~\cite{walkega2019reasoning}, amongst others \cite{DBLP:conf/edbt/NisslS22,mori-etal-2022-neural}. 
DatalogMTL extends Datalog~\cite{ceri1989you,abiteboul1995foundations} with operators from metric temporal logic~\cite{koymans1990specifying} interpreted over the rational timeline.  For example, the following rule states that travellers can enter the US if 
they had a negative test sometime  in the last $2$ days ($\diamondminus_{[0,2]}$) and have held fully vaccinated status throughout the last $15$ days ($\boxminus_{[0,15]}$):
\begin{align*}
\mathit{Authorised(x)}\gets \diamondminus_{[0,2]}\mathit{NegativeLFT(x)} \land \boxminus_{[0,15]}\mathit{FullyVaccinated(x)}.
\end{align*}
Datasets in this setting consist of temporal facts composed of 
a first-order fact annotated with a temporal interval, for example, $Authorised(John)@[13, 213.5]$. 

DatalogMTL is a powerful KR language and standard reasoning tasks, such as consistency and fact entailment, are \textsc{PSpace}-complete in data complexity \cite{walega2019datalogmtl}. This makes efficient implementation in data-intensive applications challenging. 

The most common technique of choice in scalable Datalog reasoners is materialisation (a.k.a., forward chaining) \cite{DBLP:conf/rweb/BryEEFGLLPW07,motik2014parallel,DBLP:conf/semweb/CarralDGJKU19,DBLP:journals/pvldb/BellomariniSG18}.
Facts entailed by a program and dataset are derived in successive rounds of rule applications
until  a fixpoint is reached; both this process and its output are often referred to as \emph{materialisation}. The semina\"ive algorithm forms the basis for efficient implementation by ensuring that each inference during rule application is only performed once,  thus eliminating redundant computations. 
%Materialisations can be large, but they can usually be handled on modern hardware as the available memory is continually increasing. Furthermore, 
Once the materialisation has been computed, queries can be answered directly and rules are not further considered. 

The use of metric temporal operators in rules, however, introduces a number of challenges for materialisation-based reasoning.
First, interpretations over the rational timeline are intrinsically infinite, whereas partial materialisations computed during reasoning must be finitely represented. Second, 
in contrast to Datalog where materialisation naturally terminates, in DatalogMTL a fixpoint may only be reachable after infinitely many rounds of rule applications. As a matter of fact, reasoning techniques initially proposed for \MTL{} were not materialisation-based. In particular, optimal decision procedures are automata-based \cite{walega2019datalogmtl}, and reasoning is also feasible by reduction to satisfiability checking in linear temporal logic (LTL) \cite{brandt2018querying}; finally, the Ontop system 
implements a  query rewriting approach which is applicable only to non-recursive programs \cite{guzel2018ontop}.

In our recent work \cite{wang2022meteor}, we proposed a materialisation-based procedure optimised for efficient application of DatalogMTL rules by means of suitable temporal indices, and where  partial materialisations are succinctly represented as sets of temporal facts. 
We also identified a fragment of DatalogMTL \cite{walega2021finitely} for which our materialisation-based procedure is guaranteed
to terminate; this fragment imposes suitable  
restrictions  which effectively disallow
programs expressing `recursion through time'. 
To ensure termination in the general case  for consistency and fact entailment tasks, we proposed and implemented in the MeTeoR system \cite{wang2022meteor}  an algorithm  combining materialisation with  the construction of B\"uchi automata,  so that the use of automata-based techniques is minimised  in favour of  materialisation; thus, the scalability of this approach in most practical cases is critically dependent on that of its materialisation component. The materialisation-based procedure in MeTeoR is, however, based on a \emph{na\"ive}  strategy where the main source of inefficiency stems from redundant computations.

In this paper, we propose a semina\"ive materialisation-based procedure for DatalogMTL, which can be seamlessly applied in isolation to finitely materialisable fragments \cite{walega2021finitely}, or used in the general case in combination with automata-based techniques \cite{wang2022meteor}.  As in  \cite{wang2022meteor}, our procedure iteratively performs \emph{materialisation steps} which compute partial materialisations consisting of temporal facts; furthermore, each materialisation step performs a round of rule applications followed by a coalescing phase where temporal facts differing only in their (overlapping) intervals are merged together. However, in contrast to \cite{wang2022meteor} and analogously to the classical semina\"ive algorithm for Datalog~\cite{abiteboul1995foundations},  our procedure aims at minimising redundant computation by considering only rule instances that involve information newly derived in the previous materialisation step. Lifting the  semina\"ive strategy to DatalogMTL involves significant technical challenges.
In particular, rule bodies now involve metric atoms, and derived temporal facts can be coalesced with existing facts in the previous partial materialisation.
As a result, keeping track of new information and identifying the relevant rule instances to consider in each materialisation step becomes much more involved than in Datalog.
We show that these difficulties can be overcome in an elegant and effective way, and propose additional optimisations aimed at further reducing redundancy for programs satisfying certain syntactic restrictions. 

We have implemented our approach as an extension of MeTeoR and evaluated its performance. Our experiments show that our semina\"ive strategy and the additional optimisations lead to significant reductions in materialisation times.
A technical appendix containing full proofs of our technical results and the code for our implementation are available online.\footnote{\url{https://github.com/wdimmy/MeTeoR/tree/main/experiments/RR2022}.} 

\section{Preliminaries}

We recapitulate the
definition of \MTL{} interpreted under the standard continuous
semantics for the rational timeline \cite{brandt2018querying}.

A \emph{relational atom} is a function-free first-order atom of the form $P(\sbf)$,
with $P$ a predicate and $\sbf$ a tuple of terms.
A \emph{metric atom} is an expression
given by the following grammar, where $P(\sbf)$ is a relational atom, and
$\diamondminus$, $\diamondplus$, $\boxminus$, $\boxplus$, $\Si$, $\Ui$ are MTL
operators indexed with intervals $\varrho$ containing only non-negative rationals:
\begin{equation*}
	M  \Coloneqq
		\top  \mid  \bot  \mid
	P(\sbf)  \mid
	\diamondminus_\varrho M   \mid
	\diamondplus_\varrho M   \mid   
	\boxminus_\varrho M   \mid
	\boxplus_\varrho M  \mid
	M \Si_\varrho M   \mid
	M \Ui_\varrho M.
\end{equation*}
% In particular, for a given \emph{metric atom} $M$, we use $M^p$ to denote its corresponding relational atom\nb{firs: there can be many such atoms, and second, this seems to be what you have in Definition 1.} with no MLT\nb{typo} operator and   $M^{m}$ to denote the set of MTL operators $M$ has. If $M$ is a relational atom, then  $M^p=M$\nb{should be $\{ M\}$ and not $M$} and $M^{m}=\emptyset$.
We  call $\diamondminus$, $\boxminus$, and $\Si$ \emph{past operators} and we call $\diamondplus$, $\boxplus$, and $\Ui$ \emph{future operators}. 
A \emph{rule} is an expression of the form 
\begin{equation}
M'  \leftarrow \matA_1 \land \dots \land \matA_n, \quad  \text{ for } n \geq 1, \label{eq:rule}
\end{equation}
with each $\matA_i$ a metric atom, and $\matA'$ is  generated by the following grammar:\footnote{For presentation convenience, we disallow $\bot$ in rule heads, which ensures satisfiability and allows us to focus on the materialisation process itself. } 
%In particular, we call $\diamondminus$, $\boxminus$, and $\Si$ \emph{past operators} and we call $\diamondplus$, $\boxplus$, and $\Ui$ \emph{future operators}. 
%
\begin{equation*}%\label{eq:head}
	\matA'   \Coloneqq
  	\top \mid P(\sbf)  \mid
	\boxminus_\varrho M'   \mid
	\boxplus_\varrho M'. 
\end{equation*}
The conjunction $\matA_1 \land
\dots \land \matA_n$ in Expression~\eqref{eq:rule} is the rule's \emph{body} and $\matA'$ is the rule's \emph{head}. 
A rule is \emph{forward-propagating} if it  does not mention $\top$ or $\bot$,  mentions only past operators in the body, and only future operators  in the head.
A rule is \emph{backwards-propagating}, if it satisfies analogous conditions but with past operators replaced with  future operators and vice versa.
A rule is \emph{safe} if each variable
in its head also occurs in the body, and this occurrence is not in a
left operand of $\Si$ or $\Ui$.
A \emph{program}
is a finite set of safe rules; 
it is forward- or backward-propagating  if so are all its rules.

An expression  is \emph{ground} if it mentions no variables.
A \emph{fact} is an expression $\matA @ \varrho$ with $\matA$ a
ground relational atom and $\varrho$ an interval; a
\emph{dataset}  is a finite set of facts.
The \emph{coalescing}
of facts $M@\varrho_1$ and $M@\varrho_2$, where $\varrho_1$ and $\varrho_2$ are  adjacent or have
a non-empty intersection,
is the fact $M@\varrho_3$ with $\varrho_3$ the union of $\varrho_1$
and $\varrho_2$. 
The \emph{grounding} 
$\ground{\Prog}{\D}$ 
of program $\Prog$ with respect to dataset $\D$ is the set of ground rules  obtained 
by assigning constants in $\Prog$ or $\D$  to variables in  $\Prog$.

An \emph{interpretation} $\I$ specifies, for each ground relational atom $M$ and each time point $t \in \mathbb Q$, whether $M$
holds at $t$, in which case we write ${\mathfrak{I},t \models M}$.
This extends to atoms with metric operators as shown in Table~\ref{semantics}.
\begin{table}[t]
\begin{alignat*}{3}
	&\I, t \models  \top    && && \text{for each } t
	\\
	&\I, t  \models \bot   && && \text{for no } t
	\\
	&\I,  t  \models \diamondminus_\varrho M    && \text{iff}   && \I, t' \models  M \text{ for some } t' \text{ with } t -   t' \in \varrho
	\\
	&\I,  t  \models  \diamondplus_\varrho  M  && \text{iff} &&  \I,  t' \models    M  \text{ for some } t'  \text{ with } t' - t \in \varrho
	\\
	&\I, t  \models \boxminus_\varrho  M  && \text{iff} && \I, t' \models M \text{ for all } t' \text{ with } t-t' \in \varrho
	\\
	&\I,t  \models \boxplus_\varrho  M   && \text{iff} && \I,t' \models  M \text{ for all } t'  \text{ with } t'-t \in \varrho
	\\
	&\I,t  \models M_1 \Si_\varrho M_2   &&  \text{iff} &&  \I,t' \models  M_2 \text{ for some } t'  \text{ with } t-t' \in \varrho \text{ and}
	\\
	& && &&   \I,t'' \models  M_1 \text{ for all } t'' \in (t',t)
	\\
	&\I,t  \models  M_1  \Ui_\varrho M_2 \qquad   &&  \text{iff} \qquad && \I, t' \models  M_2 \text{ for some } t'  \text{ with } t' - t \in \varrho \text{ and}
	\\
	& && &&
	  \I, t'' \models  M_1 \text{ for all } t'' \in (t,t')
\end{alignat*}
\caption{Semantics of ground metric atoms}
\label{semantics}
\end{table}
For an interpretation $\I$ and an interval $\varrho$, we define the \emph{projection}
%\nb{P: If we do not use projections in the paper, then we should delete this definition.} 
$\I\mid_{\varrho}$ of $\I$ over $\varrho$ as the interpretation that coincides with $\I$ on $\varrho$ and makes all relational atoms false outside $\varrho$. 
An interpretation $\I$ satisfies a  fact  $\matA @ \varrho$
if $\I,t \models \matA$ for all $t \in \varrho$. Interpretation $\I$ satisfies a ground rule $r$ if, whenever $\I$ satisfies each body atom of $r$ at a time point $t$, then $\I$ also satisfies the head of $r$ at $t$.
Interpretation $\I$ satisfies a (non-ground) rule $r$ if it satisfies each ground  instance of~$r$. Interpretation $\I$ is a \emph{model} of a program $\Prog$ if
it satisfies each rule in $\Prog$, and it is a \emph{model} of a dataset $\D$ if
it satisfies each fact in $\D$.
Program $\Prog$ and dataset $\D$
are \emph{consistent} if they have a model, and they
\emph{entail} a fact $M@ \varrho$
if each  model of both $\Prog$ and $\D$ is  a model of $M@ \varrho$.
Each dataset $\D$ has a unique least model $\I_\D$, and we say that dataset $\D$ \emph{represents}  interpretation $\I_{\D}$.

The \emph{immediate consequence operator} $T_{\Prog}$ for a
program $\Prog$ is a function mapping an
interpretation $\I$  to the least interpretation  containing
$\I$ and satisfying the following property for each
ground instance $r$ of a rule in $\Prog$: whenever 
$\I$ satisfies each body atom of $r$ at time point $t$, then
$T_{\Prog}(\I)$ satisfies the head of $r$ at $t$.
The successive application of $T_{\Prog}$ to $\I_\D$ defines
a transfinite sequence of interpretations $T_{\Prog}^{\alpha}(\I_\D)$ for ordinals~$\alpha$
as follows: (i)~${T_{\Prog}^0(\I_\D) = \I_\D}$, (ii)~${T_{\Prog}^{\alpha+1}(\I_\D) = T_{\Prog}(T_{\Prog}^{\alpha}(\I_\D))}$ for $\alpha$ an ordinal, and
(iii)~$T_{\Prog}^{\alpha} (\I_\D) = \bigcup_{\beta < \alpha} T_{\Prog}^{\beta}(\I_\D)$ for
$\alpha$ a limit ordinal.
The \emph{canonical interpretation} $\can{\Prog}{\D}$ of $\Prog$  and $\D$ is the interpretation
$T_{\Prog}^{\omega_1}(\I_\D)$, with
$\omega_1$ the first uncountable
ordinal.
If $\Prog$ and $\D$ have a model,
the canonical interpretation $\can{\Prog}{\D}$ 
is the least model of $\Prog$ and $\D$  \cite{brandt2018querying}.
\section{Na\"ive Materialisation in DatalogMTL}\label{sec:motivation}

In this section, we formulate the na\"ive materialisation procedure implicit in our theoretical results in \cite{walega2021finitely} and implemented in the MeTeoR reasoner \cite{wang2022meteor}.

In order to illustrate the execution of the algorithm and 
discuss the inefficiencies involved, let us consider as a running example the dataset 
$$\D_{\mathsf{ex}}=\{R_1(c_1,c_2)@[0,1], R_2(c_1, c_2)@[1,2], R_3(c_2,c_3)@[2,3], R_5(c_2)@[0,1] \}$$
and the program $\Prog_{\mathsf{ex}}$ consisting of the 
following rules:
\begin{align} 
R_1(x,y) &\gets \diamondminus_{[1,1]} R_1(x,y)    \subtag{$r_1$} ,\\ 
\boxplus_{[1,1]}R_5(y)&\gets R_2(x,y) \land \boxplus_{[1,2]}R_3(y,z) \subtag{$r_2$}, \\
R_4(x)&\gets \diamondminus_{[0,1]}R_5(x) \subtag{$r_3$}, \\
R_6(y)&\gets R_5(y) \land   \boxminus_{[0,2]}R_4(y)\land R_1(x,y).\subtag{$r_4$}
\end{align}

The na\"ive materialisation procedure applies a rule by  first identifying the facts that can ground the rule body, and then determining the maximal intervals for which all the ground body atoms hold simultaneously. 
For instance,  the procedure applies
rule $r_2$ to $\D_{\mathsf{ex}}$  by first noticing that 
relational atoms $R_2(c_1, c_2)$ and $R_3(c_2,c_3)$ can be used
to ground the rule body and then establishing that $[1,1]$ is the maximal interval for which the metric atoms $R_2(c_1, c_2)$ and $\boxplus_{[1,2]} R_3(c_2,c_3)$ in the body of the relevant instance of $r_2$ are simultaneously true in $\D_{\mathsf{ex}}$; 
as a result, $\boxplus_{[1,1]} R_5(c_2)@[1,1]$ can be derived, and so fact $ R_5(c_2)@[2,2]$
is added to the materialisation. 
In this way, the first round of  rule application of the na\"ive materialisation procedure on $\Prog_{\mathsf{ex}}$ and $\D_{\mathsf{ex}}$
also derives fact $R_1(c_1,c_2)@[1,2]$ using rule $r_1$ and fact $R_4(c_2)@[0,2]$ using $r_3$. 
The following set of facts is thus derived as a result of a single step of application of the rules in $\Prog_{\mathsf{ex}}$ to $\D_{\mathsf{ex}}$:
$$\Prog_{\mathsf{ex}}[\D_{\mathsf{ex}}] = \{R_1(c_1,c_2)@[1,2], \quad R_4(c_2)@[0,2], \quad R_5(c_2)@[2,2]\}.$$
The following definition formalises the notion of rule application.

\begin{definition}\label{def::inst}
Let $r$ be a rule of the form 
$M'  \leftarrow M_1 \land \dots \land M_n$,
for some  $n \geq 1$,
and let $\D$ be a dataset.
\noindent The set of \emph{instances} for $r$ and $\D$ is defined  as follows:
\begin{multline*}
   \ins{r}{\D}  =  \big\{ ( M_1\sigma@\varrho_1 , \dots , M_n \sigma@\varrho_n ) \mid  \sigma \text{ is a substitution and, for each } \\  i \in \{1, \dots , n\},  
   \varrho_i \text{ is a subset-maximal interval such that  } \D \models M_i\sigma @ \varrho_i \big\}.
   %\label{eq:instance}
\end{multline*}   
The set $\der{r}{\D}$ of facts \emph{derived} by $r$ from $\D$ is defined as follows:
\begin{multline}
    \der{r}{\D} = \{M \sigma@ \varrho \mid \sigma \text{ is a substitution, } M \text{ is the single relational atom in } M' \sigma,
    \\\text{and there exists } ( M_1 \sigma@\varrho_1 , \dots , M_n \sigma@\varrho_n) \in \ins{r}{\D}  \text{ such that } \varrho \text{ is the unique}
    \\
      \text{ subset-maximal  interval satisfying }  M' \sigma@(\varrho_1 \cap \ldots \cap \varrho_n) \models M\sigma@ \varrho \}.
\label{eq:consequences}
\end{multline}
The set of facts \emph{derived} from $\D$ by one-step application of  $\Prog$  is
\begin{equation}\label{eq:prog-cons}
\Prog[\D] = \bigcup_{r \in \Prog} \der{r}{\D}.
\end{equation}   
\end{definition}

Once rule application has been completed and facts
$\Prog_{\mathsf{ex}}[\D_{\mathsf{ex}}]$ have been derived, the partial materialisation $\D_{\mathsf{ex}}^1$ that will be passed on to the next materialisation step is
obtained as $\D_{\mathsf{ex}}^1 = \D_{\mathsf{ex}} \doublecup \Prog_{\mathsf{ex}}[\D_{\mathsf{ex}}]$ by coalescing facts in $\D_\mathsf{ex}$ and $\Prog_{\mathsf{ex}}[\D_{\mathsf{ex}}]$, where the coalescing operator $\doublecup$ is semantically defined next.

\begin{definition}\label{def:coalesce}
For datasets $\D_1$ and $\D_2$, we define
$\D_1 \doublecup \D_2$ as the dataset consisting of all relational facts
$M@\varrho$ such that 
$\D_1 \cup \D_2 \models M@\varrho$ and
$\D_1 \cup \D_2 \not\models M@\varrho'$, for
each $\varrho'$ 
with $\varrho \subsetneq \varrho'$.
\end{definition}

The use of coalescing makes sure that intervals associated to facts are maximal.
In our example, facts $R_1(c_1,c_2)@[0,1]$ in $\D_{\mathsf{ex}}$ and $R_1(c_1,c_2)@[1,2]$ are coalesced, so we have $R_1(c_1,c_2)@[0,2]$
in $\D_{\mathsf{ex}} \doublecup \Prog_{\mathsf{ex}}[\D_{\mathsf{ex}}]$. Thus, 
\begin{multline*}
\D_{\mathsf{ex}}^1 = \{R_1(c_1,c_2)@[0,2], R_2(c_1, c_2)@[1,2], R_3(c_2,c_3)@[2,3], \\ R_5(c_2)@[0,1], R_4(c_2)@[0,2], R_5(c_2)@[2,2]\}.
\end{multline*}

In the second round, rules are applied to $\D_{\mathsf{ex}}^1$.
The application of $r_1$ derives fact $R_1(c_1,c_2)@[1,3]$ (from $R_1(c_1, c_2)@[0,2]$) and the application of $r_2$ rederives a redundant fact $R_5(c_2)@[2,2]$. In contrast to the previous step, rule $r_4$ can now be applied to derive 
the new fact $R_6(c_2)@[2,2]$. Finally, the application of $r_3$ derives the new fact $R_4(c_2)[2,3]$ and rederives the redundant fact $R_4(c_2)[0,2]$.

The procedure then coalesces fact $R_1(c_1,c_2)[1,3]$ with $R_1(c_1,c_2)[0,2]$ to obtain $R_1(c_1,c_2)[0,3]$; similarly,  $R_4(c_2)[2,3]$ is coalesced with $R_4(c_2)[0,2]$ to obtain $R_4(c_2)[0,3]$.
Thus, the second step yields the 
following partial materialisation:
\begin{multline*}
\D_{\mathsf{ex}}^2 = (\D_{\mathsf{ex}^1} \setminus \{R_4(c_2)[0,2],R_1(c_1,c_2)[0,2]\} ) \cup {} 
\\ 
\{ R_1(c_1,c_2)@[0,3], R_6(c_2)@[2,2], R_4(c_2)[0,3]  \}.
\end{multline*}
In the third materialisation step,
rules are applied to $\D_{\mathsf{ex}}^2$, and derive
the new fact $R_1(c_1,c_2)@[1,4]$, as well
as redundant facts such as $R_5(c_2)@[2,2]$, $R_4(c_2)@[0,2]$, $R_4(c_2)@[2,3]$, and $R_6(c_2)@[2,2]$. The procedure will then continue completing subsequent materialisation steps and stopping only if a fixpoint is reached. 

\begin{algorithm}[t]
\SetKwFor{Loop}{loop}{}{}
\SetKwInput{Input}{Input}
\SetKwInput{Output}{Output}
\SetKwComment{Comment}{$//$ }{}
\Input{A program $\Prog$ and a dataset $\D$}
\Output{A dataset representing the canonical interpretation $\can{\Prog}{\D}$}

Initialise $\mathcal{N}$ to $\emptyset$ and $\D'$ to $\D$;\label{naive_init}

\Loop{}{\label{line2}
 
  $\mathcal{N} \coloneqq \Prog(\D')$\Comment*{derive new facts} \label{alg_n} 
  
  $\mathcal{C} \coloneqq \D' \doublecup \mathcal{N}$\Comment*{coalesce  with the new  facts} \label{alg_coal}
  \lIf{$\mathcal{C} = \D'$}{\Return{$\D'$}} \label{alg_fixp}
  %\Comment*{if fixpoint, return}
$\D' \coloneqq \mathcal{C};$ \label{alg_passon}
}
\caption{Na\"ive($\Prog$,$\D$)}
\label{alg::naive}
\end{algorithm}

Procedure \ref{alg::naive} formalises this na\"ive materialisation strategy. As discussed, each iteration of the main loop captures a single materialisation step consisting of a round of rule application (c.f.\ Line \ref{alg_n})  followed by the coalescing of relevant facts (c.f.\ Line \ref{alg_coal}). The resulting partial materialisation passed on to the following materialisation stem is
stored as a dataset $\D'$ (c.f.\ Line \ref{alg_passon}). 
The procedure stops when a materialisation step does not derive any new facts, in which case a fixpoint has been reached (c.f.\ Line \ref{alg_fixp}). In our example, materialisation will continue to recursively propagate the relational fact $R_1(c_1,c_2)$ throughout the infinite  timeline, and the procedure will not terminate as a result. Furthermore, the number of redundant computations will  increase in each subsequent materialisation step.

It is worth recalling that, even in cases where materialisation does not reach a fixpoint, it still constitutes a key component of terminating algorithms such as that implemented in MeTeoR \cite{wang2022meteor}. Therefore, the performance challenges stemming from redundant computations remain a very significant issue in practice. 

\section{Semina\"ive Evaluation}\label{subsec:seminaive}

Semina\"ive rule evaluation is the technique of choice for eliminating redundant computations in Datalog-based systems. The main idea is to keep track of newly derived facts in each materialisation step by storing them in a set $\Delta$, and
to make sure that rule applications in the following materialisation step involve at least one fact in $\Delta$. In this way, the procedure considers each rule instance \emph{at most once} throughout its entire execution and it is said to enjoy the \emph{non-repetition} property. Note, however, that the same fact can still be derived multiple times by \emph{different} rule instances; this type of redundancy is difficult to prevent and is not addressed by the standard semina\"ive strategy. 

Our aim in this section is to lift semina\"ive rule evaluation to the setting of DatalogMTL.
 As discussed in Section \ref{sec:motivation} on our running example, a rule instance can be considered multiple times in our setting; for example, the  instance $(R_2(c_1,c_2)@[1,2], \boxplus_{[1,2]} R_3(c_2,c_3)@[1,1])$ 
 % the original is \boxplus_{[1,2]} R_3(c_2,c_3)@[1,2]
  of $r_2$ is considered in both the first and second materialisation steps to derive  $R_5(c_2)@[2,2]$ twice since the na\"ive procedure cannot detect that
facts $R_2(c_1,c_2)@[1,2]$ and $R_3(c_2,c_3)@[2,3]$ used to instantiate $r_2$ in the second step had previously 
been used to instantiate $r_2$.
Preventing such redundant computations, however, involves certain challenges.
First, by including in $\Delta$ just the newly derived facts as in Datalog, we could overlook relevant information obtained by coalescing newly derived facts with previously derived ones.
Second, restricting application to relevant rule instances requires 
taking into account the semantics of metric operators in rule bodies.

Procedure~\ref{alg::seminaive} extends the semina\"ive strategy to the setting of DatalogMTL while overcoming the aforementioned difficulties.
Analogously to the na\"ive approach, each iteration of the main loop captures a single materialisation step consisting of a round of rule applications followed by the coalescing of relevant facts; as before, dataset $\D'$ stores the partial materialisation resulting from each iteration and is initialised as the input dataset, whereas dataset $\mathcal{N}$ stores the facts obtained as a result of rule application and is initialised as empty.

\begin{algorithm}[t]
\SetKwFor{Loop}{loop}{}{}
\SetKwInput{Input}{Input}
\SetKwInput{Output}{Output}
\Input{A program $\Prog$ and a dataset $\D$}
\Output{A dataset representing the canonical interpretation $\can{\Prog}{\D}$}

Initialise $\mathcal{N}$ to $\emptyset$, and both $\Delta$ and $\D'$
to $\D$\label{alg2_init};

\Loop{}{\label{line2}
 
  $\mathcal{N} \coloneqq \semi{\Prog}{\D'}{\Delta}$; \label{alg2_5}
  
  $\mathcal{C} \coloneqq \D' \doublecup \mathcal{N}$; \label{alg2_coal}
  
  $\Delta \coloneqq \{ M@\varrho\in  \mathcal{C} \mid M@\varrho \text{ entails some fact in } \mathcal{N} \bbslash \D'\}; $\label{alg2_delta}
  
  \lIf{$\Delta = \emptyset$}{\Return{$\D'$}\label{alg2_7}}
  $\D' \coloneqq \mathcal{C}$\label{alg2_8};
}
\caption{Semina\"ive($\Prog$,$\D$)}
\label{alg::seminaive}
\end{algorithm}

Following the rationale behind the semina\"ive strategy for Datalog,
newly derived information in each materialisation step is now stored as a dataset $\Delta$, which is initialised as the input dataset $\D$ and which is suitably maintained in each iteration; furthermore,
Procedure \ref{alg::seminaive} ensures in Line \ref{alg2_5} that only rule instances, for which it is essential to involve facts from $\Delta$ (as formalised in the following definition) are taken into account during rule application. 

\begin{definition}\label{def::semi-inst}
Let $r$ be a rule of the form
$M'  \leftarrow M_1 \land \dots \land M_n$,
for some  $n \geq 1$,
and let  $\D$ and $\Delta$ be datasets.
The set of instances for $r$ and $\D$ relative to $\Delta$
is defined as follows: 
\begin{multline}
\ins{r}{\D \threedotcolon \Delta}  =  \big\{  ( M_1\sigma@\varrho_1 , \dots , M_n \sigma@\varrho_n ) \in \ins{r}{\D}  \mid
\\
 \D \setminus \Delta \not\models
%( \D \setminus \Delta ) \not\models
M_i\sigma @\varrho_i ,  \text{ for some } i \in \{1, \dots, n \}  \big\}. \label{eq:rel-instance-delta}
\end{multline}
The set $\der{r}{\D \threedotcolon  \Delta}$ of facts derived by $r$ from $\D$ relative to $\Delta$ is defined analogously to $\der{r}{\D}$ in Definition \ref{def::inst}, with the exception that  $\ins{r}{\D}$ is replaced with $\ins{r}{\D \threedotcolon \Delta}$ in Expression \eqref{eq:consequences}.
Finally, the set $\semi{\Prog}{\D}{\Delta}$ of facts derived from $\D$
by one-step semina\"ive application of $\Prog$ is defined as $\Prog[\D]$ in Expression \eqref{eq:prog-cons}, by replacing  
$\der{r}{\D}$ with $\der{r}{\D \threedotcolon \Delta}$.
\end{definition}

In each materialisation step, Procedure~\ref{alg::seminaive} exploits  \Cref{def::semi-inst} to identify as relevant the subset of rule instances where some conjunct is `new', in the sense that 
it cannot be entailed 
without the facts in $\Delta$. 
The facts derived by such relevant rule instances in each iteration are stored in set $\mathcal{N}$ (c.f.\ Line \ref{alg2_5}). 

As in the na\"ive approach, rule application is followed by a coalescing step where the partial materialisation is updated with the facts derived from rule application (c.f.\ Line \ref{alg2_coal}). In contrast to the na\"ive approach, however, Procedure~\ref{alg::seminaive} needs to maintain set $\Delta$ to ensure that it captures only 
new facts.  This is achieved in  Line \ref{alg2_delta},
where a fact in the updated partial materialisation
is considered new if it entails a fact in $\mathcal{N}$ that was not already entailed by the
previous partial materialisation.
% where a fact in the updated partial materialisation is considered new if it entails a fact that has been derived in the current materialisation step but which was not present already in the previous partial materialisation. 
This is formalised with the following notion of `semantic' difference between temporal datasets.

\begin{definition}
Let $\D_1$ and $\D_2$ be datasets. We define $\D_1 \bbslash \D_2$ as the dataset consisting of all relational facts $M@\varrho$ such that
$M@\varrho \in \D_1$ and
$\D_2 \not\models M@\varrho$.
\end{definition}

The procedure terminates in Line \ref{alg2_7} if  $\Delta$ is empty. Otherwise, the procedure carries over the updated partial materialisation and the set of newly derived facts to the next materialisation step.

We next illustrate the execution of the  procedure on $\D_{\mathsf{ex}}$ and $\Prog_{\mathsf{ex}}$.
In the first materialisation step, all input facts are considered as newly derived (i.e., $\Delta = \D$) and hence 
$\mathcal{N} = \semi{\Prog}{\D'}{\Delta} = \Prog(\D')$ and the result of coalescing coincides with the partial materialisation computed by the na\"ive procedure (i.e., $\mathcal{C} = \D_{\mathsf{ex}}^1)$.
Then, the procedure identifies as new all facts in $\mathcal{N}$ (i.e., $\Delta = \mathcal{N}$).
In the second step, rule evaluation in Line \ref{alg2_5} no longer considers
% $\diamondminus_{[1,1]} R_1(c_1,c_2)@[1,1]$ of $r_1$ since fact
% $R_1(c_1,c_2)@[1,1] \in \D' \setminus \Delta$ already entails 
% $\diamondminus_{[1,1]} R_1(c_1,c_2)@[1,1]$. Similarly, 
the redundant instance  of $r_2$ consisting of fact $R_2(c_1,c_2)@[1,2]$ and metric atom $\boxplus_{[1,2]}R_3(c_2,c_3)@[1,2]$ since they are respectively entailed by facts $R_2(c_1,c_2)[1,2]$ and $R_3(c_2,c_3)[2,3]$ in $\D' \setminus \Delta$. 
Finally, the procedure also disregards the redundant instance of $r_3$
re-deriving fact $R_4(c_2)[0,2]$.
In contrast, all non-redundant facts derived by the na\"ive strategy are also derived by the semina\"ive procedure and after coalescing dataset $\mathcal{C} = \D_{\mathsf{ex}}^2$. 
Set $\Delta$ is now updated as follows: 
$$
\Delta = \{ R_1(c_2,c_2)@[0,3], R_6(c_2)@[2,2], R_4(c_2)@[0,3]\}.
$$
In particular, note that $\Delta$ contains the coalesced fact
$R_4(c_2)@[0,3]$ rather than fact $R_4(c_2)@[2,3]$ derived from rule application. Datasets $\Delta$ and $\D' = \D_{\mathsf{ex}}^2$ are passed on to the third materialisation step, where all redundant computations identified in Section \ref{sec:motivation} are avoided with the only exception 
of fact $R_6(c_2)@[2,2]$, which is re-derived using the instance of $r_4$ consisting of facts $R_5(c_2)@[2,2]$, $R_1(c_2, c_2)@[0,3]$
and metric atom $\boxminus_{[0,2]}R_4(c_2)@[2,3]$. Note that
this is a new instance which was not used in previous iterations, and hence the non-repetition property  remains true. Note also that, 
as with the na\"ive strategy, our semina\"ive procedure does not terminate on our running example.

We conclude this section by establishing correctness of our 
procedure.
To this end  we next show that, 
upon completion of the $k$-th iteration of the main loop (for any $k$), the partial materialisation $\D'$ passed on to the next iteration represents the interpretation $T_{\Prog}^{k}(\I_\D)$ obtained by applying $k$ times the immediate consequence operator $T_{\Prog}$ for the input program $\Prog$ to the interpretation $\I_{\D}$ representing the input dataset $\D$.
This provides a precise correspondence between the procedure's syntactic operations and the semantics of fixpoint computation.

Soundness relies on the observation that rule instances processed by
semina\"ive evaluation 
are also processed by the na\"ive evaluation; thus,
$\ins{r}{\D \threedotcolon \Delta} \subseteq \ins{r}{\D}$, for each $r$, $\D$, and $\Delta$.
As a result, each fact derived by the semina\"ive evaluation is also derived by the na\"ive evaluation. 

\begin{theorem}[Soundness]\label{soundness_sm}
Consider Procedure \ref{alg::seminaive} running on input  $\Prog$ and $\D$. 
Upon the completion of the $k$th (for some $k 
\in \N$) iteration of the loop of  Procedure~\ref{alg::seminaive}, 
it holds that  $\I_{\D'} \subseteq  T_{\Prog}^{k}(\I_\D)$.
\end{theorem}

Completeness is proved by induction on the number $k$ of iterations of the main loop.
In particular, we show that if $  T_{\Prog}^{k}(\I_\D)$ satisfies a new fact $M@t$, then there must be a rule $r$ and an instance in $\ins{r}{\D \threedotcolon \Delta}$ witnessing the derivation of $M@t$; otherwise, the fact would hold already in $  T_{\Prog}^{k-1}(\I_\D)$.
Hence, each fact satisfied by $  T_{\Prog}^{k}(\I_\D)$ is derived in the $k$th iteration of our procedure.

\begin{theorem}[Completeness]\label{complete_sm}
Consider Procedure \ref{alg::seminaive} running on input  $\Prog$ and $\D$. For each $k \in \N$, upon the completion of the $k$th iteration of the loop of  Procedure~\ref{alg::seminaive},
it holds that  $  T_{\Prog}^{k}(\I_\D) \subseteq \I_{\D'} $.
%(if the procedure terminates before the $k$th iteration, we let $\D'$ be the output of the procedure).
\end{theorem}
\section{Optimised Semina\"ive Evaluation
%Detecting Irrelevant Rules
}\label{subsec:oseminaive}

Although the semina\"ive procedure enjoys the non-repetition property, it can still re-derive facts that were already obtained in previous materialisation steps, thus incurring in a potentially large number of redundant computations. In particular, as discussed in Section \ref{subsec:seminaive}, fact $R_6(c_2)@[2,2]$ is re-derived using rule $r_4$ in the third materialisation step of our running example, and it will also be re-derived in all subsequent materialisation steps (by different instances of rule $r_4$).

In this section, we present an optimised variant of our semina\"ive procedure which further reduces the number of redundant computations performed during materialisation. The main idea is to disregard rules during the execution of the procedure as soon as we can be certain that their application will never derive new facts in subsequent materialisation steps. In our example, rule $r_4$ can be discarded after the second materialisation step as its application will only continue to re-derive  fact $R_6(c_2)@[2,2]$ in each materialisation step.

To this end, we will exploit the distinction between recursive and non-recursive predicates in a program, as defined next.

\begin{definition}\label{def:nonrecursive}
The \emph{dependency graph} of program $\Prog$ is the
directed graph 
with a vertex $v_P$ for each predicate $P$ in $\Prog$
and an edge $(v_Q, v_R)$ whenever there is a rule in $\Prog$ mentioning
$Q$ in the body and $R$ in the head.
%Program $\Prog$ is \emph{recursive} if
%$G_{\Prog}$ has a cycle.
Predicate $P$ is \emph{recursive} (in $\Prog$)
if the dependency graph has 
a path containing
a cycle and ending in $v_P$; otherwise $P$ is \emph{non-recursive}.
A metric atom is non-recursive in $\Prog$ if so are 
all its predicates; otherwise it is recursive.
The (non-)recursive fragment of $\Prog$ is the subset of
rules in $\Prog$ with (non-)recursive atoms in heads.
\end{definition}

In contrast to recursive predicates, for which new facts can be derived in each materialisation step, the materialisation of non-recursive predicates will be completed after linearly many materialisation steps; from then on, the rules will no longer derive any new facts involving these predicates.

This observation can be exploited to optimise semina\"ive evaluation. Assume that the procedure has fully materialised all non-recursive predicates in the input program. At this point, we can safely discard
all non-recursive rules;  furthermore, we can also discard a recursive rule $r$ with a non-recursive body atom $M$ if the current partial materialisation does not entail any grounding of $M$ (in this case, $r$ cannot apply in further materialisation steps).
An additional optimisation applies to forward-propagating programs, where rules cannot propagate information `backwards' along the timeline; in this case, we can compute the maximal time points  for which each non-recursive body atom in $r$ may possibly hold, select the minimum $t_r$ amongst such values, and discard $r$ as soon as we can determine that the materialisation up to time point $t_r$ has been fully completed.

The materialisation of the non-recursive predicates $R_2$, $R_3$, $R_4$, and $R_5$ of our running example is complete after two materialisation steps.
Hence, at this point we can disregard rules $r_2$ and $r_3$ and focus on the recursive forward-propagating rules $r_1$ and $r_4$. Furthermore, the maximum time point at which  $R_4$ and $R_5$ can hold is $3$ and $2$, respectively, and hence $t_{r_4} = 2$; thus,  upon completion of the second materialisation step we can be certain that $R_1$ has been materialised up to $t_r$ and we can also discard $r_6$.
In subsequent materialisation steps
we can apply only rule $r_1$, thus avoiding many redundant computations.

Procedure \ref{alg::optseminaive} implements these ideas by extending  semina\"ive materialisation. In each materialisation step, the procedure checks whether all non-recursive predicates have been fully materialised (c.f.\ Line \ref{same}), in which case it removes all non-recursive rules in the input program as well as all recursive rules with an unsatisfied non-recursive body atom (c.f.\ Lines 7--10). It also  sets a flag to $1$, which activates the additional optimisation for forward propagating programs, which is applied in Lines 11--15 whenever possible.

\begin{algorithm}[t]
\SetKwFor{Loop}{loop}{}{}
\SetKwInput{Input}{Input}
\SetKwInput{Output}{Output}
\Input{A program $\Prog$ and a dataset $\D$}
\Output{A dataset representing the canonical interpretation $\can{\Prog}{\D}$}

Initialise $\mathcal{N}$ to $\emptyset$, 
$\Delta$ and $\D'$ to $\D$, $\Prog'$ to $\Prog$,  
$flag$ to $0$, and $S_r$ (for each $r \in \Prog$) to the set of body atoms in $r$ that are non-recursive in $\Prog$;

\Loop{}{\label{loopstart}
 
$\mathcal{N} \coloneqq \semi{\Prog'}{\D'}{\Delta}$; \label{line:n}

$\mathcal{C} \coloneqq \D' \doublecup \mathcal{N}$; \label{line:c}

$\Delta \coloneqq \{ M@\varrho\in  \mathcal{C} \mid M@\varrho \text{ entails some fact in } \mathcal{N} \bbslash \D'\}; $\label{alg_delta}
  
\lIf{$\Delta = \emptyset$}{\Return{$\D'$}}

\If{$flag = 0$ $\&$ $\D'$ and $\mathcal{C}$ entail the same facts with non-recursive predicates\label{same}}
{
  Set $flag$ to $1$ and $\Prog'$ to the recursive fragment of $\Prog$; \label{delprognr}
   
   \For{each $r \in \Prog'$}{
    \lIf{there is $M \in S_r$ such that $\D' \not\models M\sigma@ t$, for each substitution $\sigma$ and time point $t$  } { $\Prog' \coloneqq \Prog' \setminus \{r\}$} \label{empty}
   }
 } 

\If{$flag = 1$ and $\Prog'$ is forward propagating\label{forward}} 
{

\For{each $r \in \Prog'$ and each $M \in S_r$}{

$t_{max}^M \coloneqq$ maximum right endpoint amongst all intervals $\varrho$ satisfying $\D' \models M\sigma@\varrho$, for some  substitution $\sigma$;

$t_{r} \coloneqq $  minimum value in $\{t_{max}^M \mid M \in S_r\}$;

\lIf{
%for each $M@\varrho \in \D'$ with $\varrho$ overlapping $(-\infty, t_r]$, there is %$M@\varrho' \in \mathcal{C}$ with $\varrho' \supseteq \varrho$
% $M@\varrho \in \D'$ implies $M@\varrho' \in \mathcal{C}$, for each $M@\varrho$ with $\varrho \cap (-\infty, t_r] \neq \emptyset$ and $\varrho \subseteq \varrho'$
%
$\D'$ and $\mathcal{C}$  coincide on all facts over intervals $\varrho$ satisfying $\varrho^+ \leq t_{r}$
\label{proj}} {$\Prog' \coloneqq \Prog' \setminus \{r\}$}
}

}
$\D' \coloneqq \mathcal{C}$;\label{loopend}
}
\caption{OptimisedSemina\"ive($\Prog$,$\D$)}
\label{alg::optseminaive}
\end{algorithm}

We conclude this section by establishing correctness of our procedure. 
We first observe that, as soon as the algorithm switches the flag to $1$, we can be certain that all non-recursive predicates have been fully materialised.
%Indeed, if two consecutive partial materialisations are the same modulo non-recursive atoms, then no new non-recursive atoms will be materialised in further iterations.

\begin{lemma}\label{non-rec}
Consider Procedure~\ref{alg::optseminaive} running on input $\Prog$ and $\D$ and let $\prognr$ be the non-recursive fragment of $\Prog$.
If $flag = 1$, then $\can{\prognr}{\D} \subseteq \I_{\D'}$.
\end{lemma} 

Next, we show how we can detect if a forward-propagating program  has  completed materialisation of all facts (also with recursive predicates) up to a given time point.
Namely it suffices to check that two consecutive partial materialisations satisfy the same facts up to a given time point.
Note that our procedure checks this condition syntactically in Line~\ref{proj}.

% The following lemmas show the correctness of our additional optimisation on forward-propagating programs. Lemma \ref{lemma:fb} shows that, for a forward-propagating program, it is possible to detect whether the materialisation has been completed up to a given time point.

\begin{lemma}\label{lemma:fb}
If $\I_{\D}\mid_{(-\infty, t]} = T_{\Prog}(\I_{\D})\mid_{(-\infty, t]}$, for a forward propagating program $\Prog$, dataset $\D$, and  time point $t$,
then $\I_{\D}\mid_{(-\infty, t]} = \can{\Prog}{\D}\mid_{(-\infty, t]}$.
%Similarly, if
%$\I_{\D}\mid_{[t,\infty)} = T_{\Prog} (\I_{\D})\mid_{[t,\infty)}$
%and $\Prog$ is backward-propagating, 
%$ { \I_{\D}\mid_{[t,\infty)} = \can{\Prog}{\D}\mid_{[t,\infty)} }$. 
\end{lemma}

We can use this lemma to show that each rule discarded in Lines 10 and 15 can be safely ignored as it will have no effect in subsequent materialisation steps.

\begin{lemma}\label{lemma:invalid}
If in  Procedure \ref{alg::optseminaive} a rule $r$ is removed from $\Prog'$ in Line 10 or in Line~15, then 
$\can{\Prog'}{\D' \doublecup \mathcal{N}} = \can{\Prog'\setminus 
\{r\}}{\D'\doublecup \mathcal{N}}$.
% Consider the execution of Line 15 in Procedure \ref{alg::optseminaive}. If the condition in the \textbf{if} statement applies, then 
% $\can{\Prog'}{\D' \doublecup \mathcal{N}} = \can{\Prog'\setminus 
% \{r\}}{\D'\doublecup \mathcal{N}}$.
%Let $\D$ be a dataset.
%Consider Algorithm~\ref{alg:invalid} running on inputs $r$, $\Prog'$, $\Prog$, $\D'$, and $\mathcal{N}$  such that $r \in \Prog' \subseteq %\Prog$, 
%$\D' \doublecup \mathcal{N} = T_{\Prog'}(\I_{\D'})$,
%and 
%$ \can{\Prog_{nr}}{\D} \subseteq \I_{\D'}$,
%for $\Prog_{nr}$ the non-recursive fragment of $\Prog$.
%If the algorithm returns $True$, then $\can{\Prog'}{\D' \doublecup %\mathcal{N}} = \can{\Prog'\setminus 
%\{r\}}{\D'\doublecup \mathcal{N}}$.
\end{lemma}

Finally, using Lemmas \ref{non-rec} and \ref{lemma:invalid} together with the soundness and completeness of our semina\"ive evaluation (established in \Cref{complete_sm,soundness_sm}), we can show soundness and completeness of the optimised version of the procedure.

\begin{theorem}[Soundness and Completeness]\label{seminaive_theorem}
Consider Procedure \ref{alg::optseminaive} running on input  $\Prog$ and $\D$. For each $k \in \N$, the partial materialisation $\D'$ obtained upon completion of the $k$th iteration of the main loop represents the interpretation $T_{\Prog}^{k}(\I_{\D})$.
\end{theorem}

We conclude by observing that our optimisation for forward-propagating programs in Lines 11--15 can be modified in a straightforward way to account also for backwards-propagating programs, as these two cases are symmetric.

\section{Evaluation}\label{sec:experiments}

We have implemented Procedures \ref{alg::seminaive} and \ref{alg::optseminaive} as an extension of our  open-source MeTeoR reasoner \cite{wang2022meteor}, which so-far implemented only the na\"ive strategy from Procedure \ref{alg::naive} to perform materialisation. 

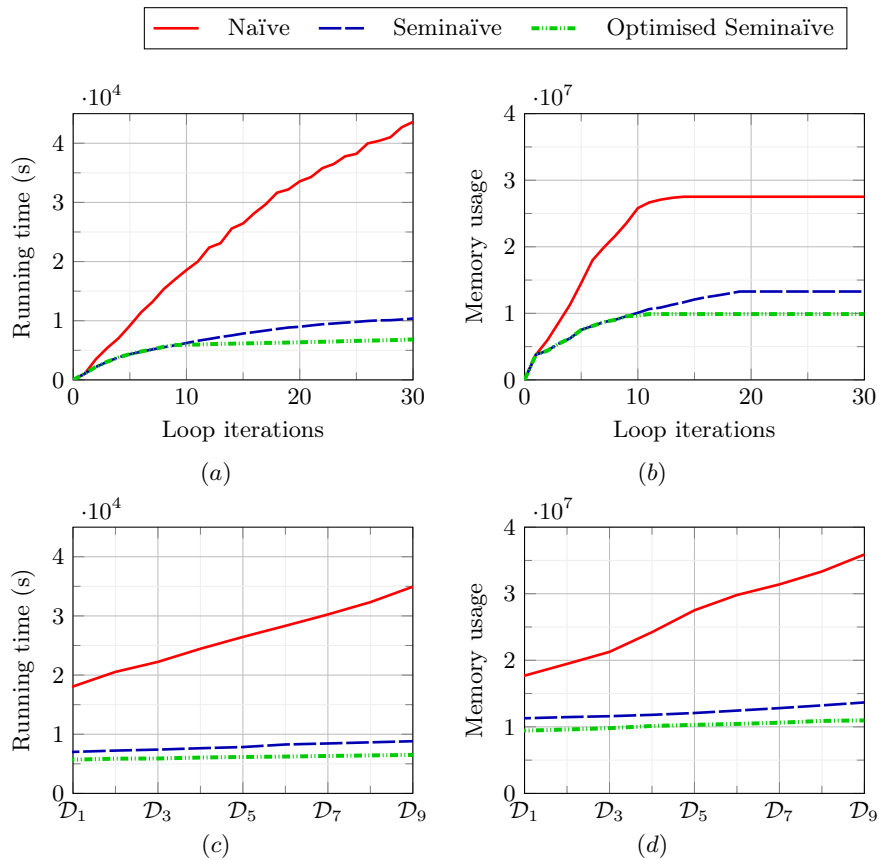
\begin{figure*}[ht]
	\centering
	\begin{tikzpicture}
		\pgfplotsset{%
			width=0.5\textwidth,
			height=0.42\textwidth
		}
		\begin{axis}[
			xshift=-6cm,
			xlabel= Loop iterations,
			ylabel=Running time~(s),
			ylabel near ticks,
			xlabel near ticks,
			xmin=0, xmax=30,
			ymin = 0, ymax=45000,
			ytick = {0, 10000, 20000,30000,40000},
			yticklabels = {0, 1, 2,3,4},
			grid = both,
            minor tick num = 1,
			major grid style = {lightgray},
            minor grid style = {lightgray!25},
			]
			\addplot[red,line width=1pt] table {tikzdata/naive_t_2.dat};
			\addplot[dash pattern=on 0.3cm off 0.05cm,black!30!blue,line width=1pt] table {tikzdata/semi_t_2.dat};
			\addplot[densely dashdotdotted,black!20!green,line width=1.5pt] table {tikzdata/opt_t_2.dat};
			
		\end{axis}
		\begin{axis}[
	     	legend style={at={(0.95,1.4)}, column sep=0.3cm, font=\small},
			legend columns=-1,
			xlabel= Loop iterations,
			ylabel= Memory usage,
			ylabel near ticks,
			xlabel near ticks,
			xmin = 0, xmax = 30,
            ymin = 0, ymax = 40000000,
            % xtick distance = 1,
            % ytick distance = 0.25,
			grid = both,
            minor tick num = 1,
			major grid style = {lightgray},
            minor grid style = {lightgray!25},
			]
			\addplot[red, line width=1pt] table {tikzdata/naive_redundancy_2.dat};
			\addplot[dash pattern=on 0.3cm off 0.05cm,black!30!blue,line width=1pt] table {tikzdata/semi_redundancy_2.dat};
			\addplot[densely dashdotdotted,black!20!green,line width=1.5pt] table {tikzdata/opt_redundancy_2.dat};
			\legend{Na\"ive, Semina\"ive, Optimised  Semina\"ive}
		\end{axis}
		
		\begin{axis}[
		    yshift=-5.5cm,
	     	%legend style={at={(2.38,1.4)}, column sep=0.3cm, font=\small},
			%legend columns=-1,
% 			xlabel=\# of relational facts,
			ylabel= Memory usage,
			ylabel near ticks,
			xlabel near ticks,
			xtick={2,4,6,8,10},
			xticklabels={$\D_{1}$, $\D_3$, $\D_5$, $\D_7$,  $\D_{9}$},
			xmin = 2, xmax = 10,
			ymin=0, ymax=40000000,
			grid = both,
            minor tick num = 1,
			major grid style = {lightgray},
            minor grid style = {lightgray!25},
			grid = both,
            minor tick num = 1,
			major grid style = {lightgray},
            minor grid style = {lightgray!25},
			]
			\addplot[red,line width=1pt] table {tikzdata/naive_redundancy_1.txt};
			\addplot[dash pattern=on 0.3cm off 0.05cm,black!30!blue,line width=1pt] table {tikzdata/semi_redundancy_1.txt};
			\addplot[densely dashdotdotted,black!20!green,line width=1.5pt] table {tikzdata/opt_redundancy_1.txt};
% 			\legend{Na\"ive, Semina\"ive, Optimised  Semina\"ive}
		\end{axis}
		\begin{axis}[
			xshift = -6cm,
		    yshift = -5.5cm,
% 			xlabel=\# of relational facts,
			ylabel=Running time~(s),
			ylabel near ticks,
			xlabel near ticks,
			xtick={2,4,6,8,10},
			xticklabels={$\D_{1}$, $\D_3$, $\D_5$, $\D_7$,  $\D_{9}$},
			ytick = {0, 10000, 20000,30000,40000},
			yticklabels = {0, 1, 2,3,4},
			xmin = 2, xmax = 10,
        	ymin = 0, ymax=45000,
			grid = both,
            minor tick num = 1,
			major grid style = {lightgray},
            minor grid style = {lightgray!25},
			]
			\addplot[red,line width=1pt] table {tikzdata/naive_t_1.txt};
			\addplot[dash pattern=on 0.3cm off 0.05cm,black!30!blue,line width=1pt] table {tikzdata/semi_t_1.txt};
			\addplot[densely dashdotdotted,black!20!green,line width=1.5pt] table {tikzdata/opt_t_1.txt};
			
		\end{axis}

		\draw (1.7,-1.25) node {$(b)$}; 
		\draw (1.7,-6.2) node {$(d)$}; 
		\draw (-4.1,-1.25) node {$(a)$}; 
		\draw (-4.1,-6.2) node {$(c)$}; 
	\end{tikzpicture}
	
	\caption{Experimental results for dataset $\D_5$ in sub-figures (a) and (b), and for the first 15 iterations for datasets $\D_1$--$\D_9$ in sub-figures (c) and (d)}
	\label{fig::comparison}
\end{figure*}

For  evaluation, we have considered the temporal extension of the Lehigh University Benchmark (LUBM) \cite{guo2005lubm} used in previous evaluations of MeTeoR  \cite{wang2022meteor}. 
The benchmark provides a DatalogMTL program  consisting
of $56$ Datalog rules obtained from the OWL 2 RL fragment of LUBM's ontology plus 
$29$ temporal rules involving recursion and covering all
metric operators of \MTL{}. 
To make materialisation more challenging, we have included additional body atoms in some of the temporal rules.
The benchmark also provides an extension of LUBM’s data generator which randomly assigns intervals non-temporal facts.
% ; the endpoints of each interval belong to a range given as a parameter. 
We used nine datasets $\D_1$--$\D_{9}$, each consisting of 10 million facts, but with an increasing number of constants occurring in these facts (and thus with a smaller number of intervals per relational fact), namely, these datasets contain 
$0.8$, $1.0$, $1.2$, $1.3$, $2.1$, $2.5$, $5.2$, $10.1$, and $15.8$
% \nb{P: Dingmin, please check these numbers.} 
million constants, respectively.
We compared running time and  memory requirements (maximal number of stored facts) of our procedures with that of the na\"ive approach as depicted in \Cref{fig::comparison}.  Experiments were conducted
on a Dell PowerEdge R730 server with 512 GB RAM and two Intel Xeon E5-2640 2.6 GHz processors running Fedora 33, kernel version 5.8.17.

Figures~\ref{fig::comparison} (a) and (b)
show time and memory usage on a single dataset $\D_5$ through the first 30 iterations of the procedures.
As we can see, the semina\"ive procedure significantly outperforms the na\"ive approach  both in terms of running time and memory consumption, especially as materialisation progresses.
In turn, the optimised semina\"ive approach is able to start disregarding rules after $9$ materialisation steps, and at that point it starts outperforming the basic semina\"ive procedure. 
We can also observe that, at this point, many of the predicates have been materialised already and the number of new facts generated in each further step is very small (thus, memory consumption stops growing). 
Despite this, the na\"ive algorithm continues to perform a large number of redundant computations, and hence its running time increases at a similar rate as before; in contrast, the optimised semina\"ive approach avoids most of this redundancy and subsequent materialisation steps are completed very efficiently.

Figures~\ref{fig::comparison}
% \nb{P: Dingmin, please check the numbers in figures, as currently they do not correspond to (a) and (b).}
(c) and (d) summarise our results for $15$
% \nb{P: Dingmin, please check this number.} 
materialisation steps on datasets $\D_1$--$\D_{9}$ with increasing numbers of constants.
We can observe that both the time and memory consumption in the na\"ive approach increase linearly (and in a significant way) with the number of constants. 
Indeed, by increasing the number of constants, we are also increasing the number of ground rule instances to be examined. 
In contrast, the effect on both of our semina\"ive procedures is much less noticeable as they can quickly disregard irrelevant instances.

% Figure~\ref{fig:intervals} summarises our results for $10$ materialisation steps on datasets $\D_i^t$ with varying number of intervals per relational atom. As we can see, by increasing the number of intervals and reducing the number of constants, memory consumption and materialisation times decrease in all approaches. First, fewer constants implies fewer ground instances to consider and hence lower running times. Second, by increasing the number of intervals the number of coalescing operations also increases and thus the total number of facts in memory tends to decrease also. Finally, MeTeoR implements a variant of  the sort-merge join algorithm which is effective at reducing temporal joins when many facts share the same relational atom, which contributes to the decrease in running times.

\section{Conclusion and Future Work}

In this paper, we have presented an optimised semina\"ive materialisation procedure for \MTL{}, which 
can efficiently materialise complex recursive programs and large datasets involving millions of temporal facts. 

We see many exciting avenues for future research. First, \MTL{} has been 
extended with 
stratified negation-as-failure
\cite{DBLP:conf/aaai/CucalaWGK21} and  our semina\"ive
procedure could be extended accordingly. It would also be interesting to consider semina\"ive evaluation for reasoning under alternative semantics for \MTL{} such as the integer semantics \cite{DBLP:conf/kr/WalegaGKK20} or the point-wise semantics \cite{ryzhikov2019data}.
We are also working on blocking conditions that exploit the periodic structure of canonical models to ensure termination of materialisation-based reasoning.
Finally, incremental materialisation-based reasoning has been studied in context of Datalog \cite{DBLP:journals/ai/MotikNPH19}, and it would be interesting to lift such approaches to the \MTL{} setting.

\section*{Acknowledgments}

This work was supported by
the EPSRC project OASIS (EP/S032347/1),
the EPSRC project UK FIRES (EP/S019111/1),
and the SIRIUS Centre for Scalable Data Access, and
Samsung Research UK.

% \nb{P: Please correct reference: e.g., LUBM, OWL, MTL, Datalog --- all with big letters}

\bibliography{reference}
\bibliographystyle{splncs04}
%\clearpage
%\appendix
%\input{appendix}

\end{document}